\documentclass[a4paper,11pt]{article}
\usepackage{pos}

\title{Hadronisation in event generators from small to large systems}

\author*[a,b,c]{Yuuka Kanakubo}
\author[c,d]{Yasuki Tachibana}
\author[c]{Tetsufumi Hirano}

\affiliation[a]{University of Jyväskylä, Department of Physics, P.O. Box 35, FI-40014 University of Jyväskylä, Finland}

\affiliation[b]{Helsinki Institute of Physics, P.O. Box 64, FI-00014 University of Helsinki, Finland}
\affiliation[c]{Department of Physics, Sophia University, Tokyo 102-8554, Japan}
\affiliation[d]{Akita International University, Yuwa, Akita-city 010-1292, Japan}

\emailAdd{yuka.y.kanakubo@jyu.fi}
\emailAdd{ytachibana@aiu.ac.jp}
\emailAdd{hirano@sophia.ac.jp}

\abstract{
The results of the dynamical core-corona initialisation framework in p+p and Pb+Pb collisions at the LHC
energies are presented. We extract the fractions of final hadron yields originating from equilibrated
and non-equilibrated matter as functions of multiplicity. We show that the contribution from non-equilibrated matter is non-negligible even in intermediate and central Pb+Pb collisions. The particle production from non-equilibrated matter behaves as a correction on $c_2\{4\}$ that is purely obtained from the equilibrated matter.
The result poses a warning on Bayesian parameter estimation with conventional hydrodynamic models. The observed flow coefficients might need a reinterpretation with new dynamical models which incorporate the particle production from non-equilibrated matter.
}

\FullConference{The Eleventh Annual Conference on Large Hadron Collider Physics (LHCP2023)\\
 22-26 May 2023\\
 Belgrade, Serbia\\}

\begin{document}
\maketitle

\section{Introduction}
The relativistic nuclear collisions at RHIC and LHC experiments have provided the opportunities to study
the properties of quark-gluon plasma (QGP), a strongly interacting many-body system consisting of quarks and gluons under local equilibrium.
The mainstream theoretical tool to study the QGP properties has been multi-stage dynamical models based on relativistic hydrodynamics.
Recent studies have shown that Bayesian parameter estimation with multi-stage dynamical models constrains the transport coefficients of QGP
\cite{Pratt:2015zsa,Bernhard:2016tnd}.
These groundbreaking works elevated the precision of QGP studies.
Given the increasing popularity of the application of the Bayesian parameter estimation following these works, the sophistication of the existing dynamical models has become inevitable.
On the other hand, the possibility of QGP formation in small colliding systems, proton--proton or proton--ion collisions etc., has been implied by some experimental data.
The signals of QGP are observed in high-multiplicity events, for instance, enhancement of multi-strange hadron productions and long-range azimuthal correlations.
The origin of these signals is still a topic of ongoing discussion and requires further investigation
\cite{Dusling:2015gta,Nagle:2018nvi,Schenke:2021mxx}.

Motivated by the above backgrounds, we establish a multi-stage dynamical framework which aims to describe from small to large colliding systems
assuming that the emergence of QGP signals in high-multiplicity small systems originates from partial equilibration of the system.
With the established framework, we try to assess this assumption by performing model-to-data comparisons for various observables.

\section{Dynamical Core–Corona Initialisation framework (DCCI)}
To realise the multiplicity-dependent partial formation of QGP in the system, we apply the idea of the core--corona picture \cite{Werner:2007bf} -- the two-component picture with equilibrated QGP (core) and non-equilibrated matter (corona) for relativistic nuclear collisions.
In midrapidity, the density of generated matter right after a collision of nuclei is large at the centre of the overlapping regions of two nuclei, and it is small at the peripheral regions.
This non-uniformity of the generated matter provides the following picture: the generation of the core component is achieved in the dense regions and the corona component corresponds to the matter generated in peripheral regions.

The above picture is dynamically realised in our framework, 
the updated dynamical core--corona initialisation (DCCI2) \cite{Kanakubo:2021qcw,Kanakubo:2022ual}.
The dynamical initialisation of QGP fluids is performed via source terms of relativistic hydrodynamic equations. 
Under the two-component picture, the energy-momentum conservation is achieved as a whole system consisting of QGP fluids and non-equilibrated partons.
Hence, the space-time evolution of QGP fluids corresponds to the deposition of energy and momentum from non-equilibrated partons.
The core--corona picture is encoded in the energy-momentum deposition rate.
We model the energy-momentum deposition so that it is proportional to the collision rate which is evaluated with a phenomenologically-defined mean-free path.
With this model, initially produced partons with low transverse momentum, $p_T$, and/or in high-density regions contribute to generating QGP fluids.  On the other hand, those with high $p_T$ and/or in low-density regions tend to keep their initial energy and momentum.

The entire model flow of DCCI2 is outlined as follows: The phase-space information of initial partons are obtained from \textsc{Pythia}8 \cite{Sjostrand:2007gs} and \textsc{Pythia}8 Angantyr \cite{Bierlich:2016smv,Bierlich:2018xfw} for p+p and A+A collisions, respectively. 
The following evolution of the system is described with (3+1)-D ideal hydrodynamics \cite{Tachibana:2014lja} while the dynamical separation of the system into the core and the corona is performed under the dynamical core--corona initialisation that we explained above. 
The hydrodynamics incorporates the equation of state $s$95$p$-v.1.1 \cite{Huovinen:2009yb}. 
At the switching temperature from hydro to particles, the particlisation is performed with iS3D \cite{McNelis:2019auj}, a Monte-Carlo sampler of hadrons from
hydrodynamic fields. The hadronisation of surviving partons is performed via the Lund string fragmentation with \textsc{Pythia}8 constructing colour singlet strings. Finally, direct hadrons emitted both from hydro and string fragmentation are handed to \textsc{Jam} \cite{Nara:1999dz}, a hadronic transport model that performs hadronic rescatterings and resonance decays.

\section{Results}
We simulate p+p at $\sqrt{s} = 7, 13$ TeV and Pb+Pb at $\sqrt{s_{\mathrm{NN}}}=2.76$ TeV. 
In the DCCI2, initial conditions are given by \textsc{Pythia}8 and \textsc{Pythia}8 Angantyr. Therefore, 
major parameters that need to be fixed to reproduce basic observables, such as multiplicity and particle yield ratios, are the ones in the dynamical core--corona initialisation.
The parameters are fixed so that the model reasonably describes the multiplicity dependence of the yield ratios of omega baryons to charged pions ($\Omega/\pi$) in p+p and Pb+Pb collisions
reported by the ALICE Collaboration \cite{ALICE:2016fzo}.
As multiplicity increases, the system becomes core-dominated from corona-dominated, which leads to the smooth enhancement of $\Omega/\pi$ along multiplicity.

This also means that we can extract the fractions of hadronic production originating from the core and the corona from multiplicity dependence of $\Omega/\pi$.
The left panel in Fig.~1 shows the fraction of the core and the corona at midrapidity (|$\eta$| < 0.5) in p+p and Pb+Pb collisions. 
The fractions of the core and the corona show clear multiplicity scaling regardless of the different system sizes or collision energies. 
The onset of the core dominance is located at $\langle dN_{\mathrm{ch}}/d\eta \rangle \sim 20$.
It is worth mentioning that the fraction of core components is only around 50\% even
at the highest multiplicity in p+p collisions, which poses caution on the simulations of such events with pure hydrodynamic frameworks.
It should be also noted that the contribution from the corona remains non-negligible even in intermediate and central Pb+Pb collisions while it is usually regarded that the system can be described with pure hydrodynamics at low $p_T$ regions.

To investigate the effect of the corona contribution in Pb+Pb collisions, we analyse one of the important flow
observables, four-particle cumulant of charged particles $c_2\{4\}$, as a function of the number of
charged particles, $N_{\mathrm{ch}}$, in the right panel of Fig.~1.
The results from the core contribution and the inclusive hadrons show the negative values of $c_2\{4\}$ manifesting the strong multi-particle correlation in Pb+Pb collisions.
The result from the corona contribution stays zero-consistent in the entire range of $N_{\mathrm{ch}}$ because back-to-back correlation is subtracted in $c_2\{4\}$.
The main message in this figure can be read from the difference of $c_2\{4\}$ between the core contribution and the inclusive hadrons.
One sees that the absolute value of $c_2\{4\}$ solely from the core contribution is diluted just due to the existence of the corona contribution in the system.
This implies that the quantitative extraction of QGP properties from model-to-data comparisons would require new dynamical models which incorporate contributions from non-equilibrated matter.

\section{Summary}
We reported the results from the updated dynamical core–corona initialisation framework which aims to describe particle production from small to large colliding systems under a unified picture.
We found that the fractions of the core and corona contributions show clear scaling
with final state multiplicity regardless of the different system sizes or collision energies.
The system becomes core-dominated above $\langle dN_{\mathrm{ch}}/d\eta \rangle \sim 20$.
We also showed that there are non-negligible
contributions from the corona even in intermediate and central Pb+Pb collisions.
Further investigation of the corona contribution in Pb+Pb collisions was performed by analysing $c_2\{4\}$.
The comparison between the core contribution and the inclusive hadrons revealed the non-equilibrium corrections on $c_2\{4\}$ obtained from hydro contribution.
Our results suggest the importance of incorporating the non-equilibrium contribution in dynamical models 
for the quantitative determination of QGP properties from model-to-data comparisons.

\begin{figure}
    \centering
    \includegraphics[width=0.47\textwidth, bb=0 0 691 447]{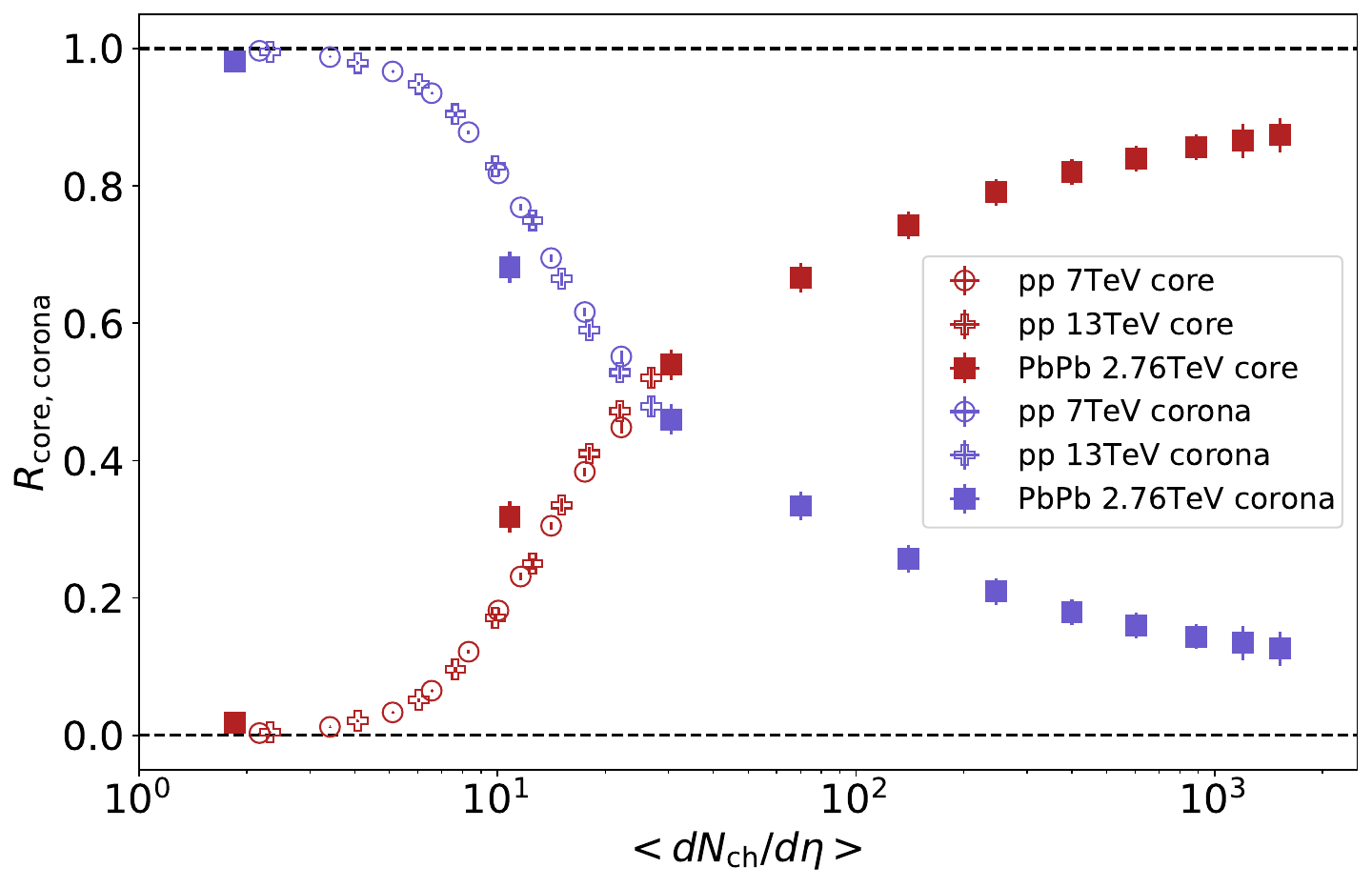}
    \hspace{5pt}
    \includegraphics[width=0.47\textwidth, bb=0 0 636 455]{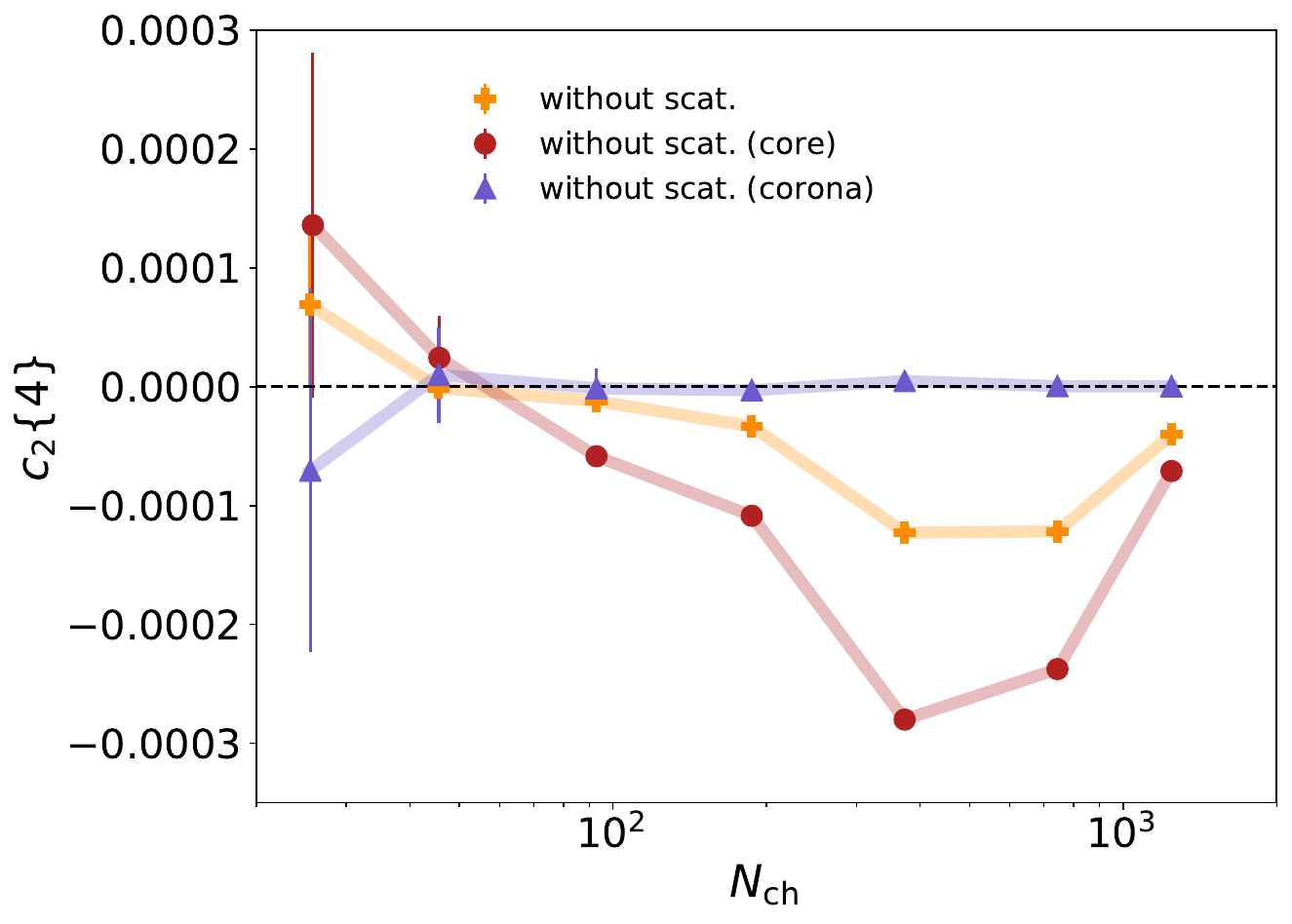}
    \caption{(Left) Fractions of hadronic production originating from the core and the corona as a function of multiplicity. Results of p+p at $\sqrt{s} = 7, 13$ TeV and 
    Pb+Pb at $\sqrt{s_{\mathrm{NN}}}=2.76$ TeV are shown. 
    (Right) Comparisons of four-particle cumulant of charged particles $c_2\{4\}$ as a function of the number of charged particles $N_{\mathrm{ch}}$ in Pb+Pb at $\sqrt{s_{\mathrm{NN}}}=2.76$ TeV
    for the core, the corona, and inclusive (core and corona) contributions.}
    \label{fig:results_DCCI}
\end{figure}

\section*{Acknowledgement}
Our research was funded as a part of the Center of Excellence in Quark Matter of
the Academy of Finland (Project No. 346325), the European Research Council Project No.
ERC-2018-ADG835105 YoctoLHC, the Academy of Finland Project No. 330448, and the
European Union’s Horizon 2020 research and innovation program under grant agreement No.
824093 (STRONG-2020).

\bibliographystyle{JHEP}
\bibliography{bibs}

\end{document}